\documentclass[
 reprint,
 amsmath,amssymb,
 aps,
prb,
 floatfix,
]{revtex4-2}
\usepackage{lipsum}
\usepackage{graphicx}
\usepackage{dcolumn}
\usepackage{bm}
\usepackage[unicode=true,
 bookmarks=false,
 breaklinks=false,pdfborder={0 0 0},pdfborderstyle={},backref=false,colorlinks=true]
 {hyperref}
\hypersetup{colorlinks,citecolor=blue,linkcolor=blue,urlcolor=blue}

\usepackage{xcolor}
\usepackage{soul}

\begin{document}
\title{Quantum circuit refrigerator based on quantum dots coupled to normal-metal and superconducting electrodes}
\author{S. Mojtaba Tabatabaei}
\affiliation{Department of Physics, Kharazmi University, Tehran, Iran}
\author{Neda Jahangiri}
\affiliation{Department of Physics, Sharif University of Technology, Tehran, Iran}
\date{\today}
\begin{abstract}
In quantum dot junctions capacitively coupled to a resonator, electron tunneling through the quantum dot can be used to transfer heat between different parts of the system. This includes cooling or heating the electrons in electrodes and absorbing or emitting photons in the resonator mode.
Such systems can be driven into a nonequilibrium state by applying either a voltage bias or a temperature gradient across the electrodes coupled to the quantum dot, or by employing an external coherent pump to excite the resonator.
In this study, we present a semiclassical theory to describe the steady state of these structures.
We employ a combination of the Floquet-nonequilibrium Green's functions method and semiclassical laser theory to analyze a normal metal-quantum dot-superconductor junction coupled to a resonator.
Our investigation focuses on key parameters such as the average photon number and phase shift in the resonator, the charge current in the quantum dot, and the heat fluxes among different components of the system.
We explore how photon-assisted Andreev reflection and quasiparticle tunneling in the quantum dot can refrigerate the resonator mode and the normal metal electrode. We also examine the influence of finite voltage and thermal biases on these processes.
\end{abstract}
\maketitle

\section{\label{sec:level1}INTRODUCTION}

In recent years, the manipulation of heat flows between different parts of quantum circuits has sparked renewed interest because of its wide-ranging applications, such as cooling quantum circuits and qubit initialization processes.\cite{Giazotto-2012,partanen2016quantum,sevriuk2019fast,maillet2020electric,elouard2020quantifying,RevModPhys.93.025005}.
Various proposals for quantum refrigerators\cite{Valenzuela-2006,PhysRevB.76.174523,PhysRevApplied.17.064022}, heat rectifiers\cite{senior2020heat,majland2020quantum,liu2023quantum}, and heat engines\cite{PhysRevLett.112.076803,Marchegiani-2016,hardal2017quantum,kamimura2022quantum} have been put forth in the circuit quantum electrodynamics (circuit QED) architecture. 

Quantum circuit cooling refers to a process aimed at reducing the effective temperature of a quantum circuit.
While quantum circuits themselves do not have a temperature in the classical sense, cooling in this context typically means bringing the system closer to a state with lower energy.
For example, it has been shown that a voltage-biased normal metal-insulator-superconductor (NIS) junction can cool the normal metal due to the Peltier effect\cite{Nahum-1994,Leivo-1996,Lowell-2013,Courtois-2014,PhysRevApplied.4.024006,Courtois-2016}. 
Simultaneously, tunneling quasiparticles can also cool a nearby resonator capacitively coupled to the junction by absorbing photons from the resonator\cite{Silveri-2019,Masuda-2018,Tan-2017,Silveri-2017,vadimov2022single,andp.202100543}.
This effect is due to the existence of an energy gap in the density of states for the superconductor, where quasiparticles with higher energy are more effectively removed from the normal metal than those with lower energy within the gap\cite{RevModPhys.78.217}.

As well, solid-state quantum dot (QD) structures connected to normal metal or superconductor electrodes have also been widely used as a platform for controlling heat flow in quantum circuits \cite{Liu-2013,Campisi-2015,mitchison2016realising,PhysRevB.98.241414,jaliel2019experimental,harvey2020chip,manikandan2020autonomous,PhysRevLett.125.247701,PhysRevB.106.115419,PhysRevResearch.5.043041,thierschmann2015three,PhysRevLett.114.146805}. In these circuits, the QDs can be conveniently coupled to coplanar waveguide (CPW) resonators \cite{Goeppl2008} through capacitive interactions between the electric charges on the QD and the electric field in the resonator\cite{Delbecq2011,Bruhat-2016,Bruhat2018}. The photon-assisted electron tunneling in the QD-resonator coupled system has also been actively studied in recent years\cite{Childress-2004,Meyer-2007,Frey-2012,Jin-2013,Liu-2014,Mavalankar-2016,PhysRevB.101.115135,Ghirri-2020,Jahangiri-2020}. It has been shown that energy exchange with the resonator can enhance the charge current through the QD, allowing the QD to operate as an efficient quantum heat engine \cite{PhysRevLett.112.076803}. Interestingly, this behavior is expected even for a QD system in equilibrium \cite{sothmann2014thermoelectric}. The physics behind these heat exchanges becomes more exciting when at least one of the electrodes connected to the QD is a superconductor.
This allows for heat exchange enhancement due to the photon-assisted Andreev reflections between the normal metal and the superconductor electrode\cite{PhysRevResearch.1.033098}. 

A useful tool for assessing the resonator cooling is the average photon number in the resonator. Thus, by assuming that the quantum circuit is initially populated with a finite average photon number, circuit refrigeration corresponds to the reduction in the average photon number of the circuit \cite{Hauss2008}. In practice, the average photon number in a CPW resonator can be determined by measuring the transmission or reflection coefficients of the resonator in response to an incoming microwave pulse from an external probe \cite{Fink2010}.

In this work, we study a quantum circuit refrigerator based on a hybrid quantum dot (QD) that is coupled to a microwave resonator. The hybrid QD subsystem consists of a single-level QD connected to a normal metal and a superconducting electrode. We study the photon absorption and emission in the resonator and the thermoelectric behavior of the QD when either a finite thermal gradient or a finite voltage bias is applied across the two electrodes connected to the QD. To this end, we determine the average photon number in the resonator by solving the Heisenberg-Langevin equation in the semiclassical approximation. Moreover, we calculate the nonequilibrium steady state of the QD when it is coupled to both the electrodes and the resonator by employing the Floquet-nonequilibrium Green's function(NEGF) method. Our results show that the QD system can act as a circuit refrigerator or as an active medium that emits photons into the resonator. These behaviors can be controlled by changing the gate voltage on the QD as well as by applying a finite electric voltage or thermal bias on the QD.

The rest of this work is structured as follows: In Sec. \ref{sec:level2}, we present the model Hamiltonian for the system. In Sections \ref{sec:level3} and \ref{sec:NEGF}, we derive the equation of motion for the photon mode in the resonator and present the Floquet-NEGF formalism for the QD subsystem, allowing us to obtain a system of equations describing the steady state of the QD-resonator coupled system. We then present our numerical results regarding cooling and photon emissions in the resonator in Sec. \ref{sec:level4}. The charge and heat currents through the QD system, as well as the cooling power of the resonator, are studied in Sec. \ref{sec:level5}. Finally, we draw our conclusions in Sec. \ref{sec:level6}.

\section{Theoretical formalism}

\subsection{\label{sec:level2}The Model Hamiltonian}

\begin{figure}
\centering{}\includegraphics[width=8cm]{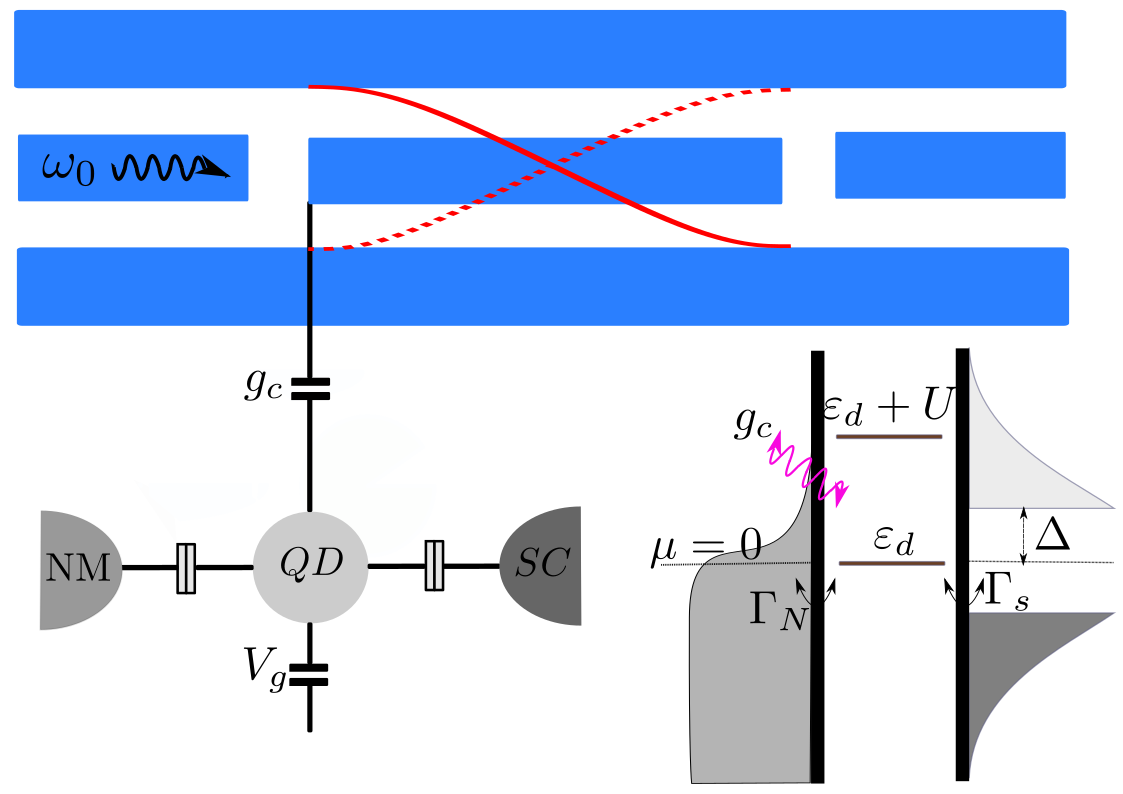} \caption{\label{fig:taba301}A single quantum dot (QD), which is coupled to a normal metal (NM) and a superconductor (SC) with an energy gap $\Delta$, is capacitively coupled to a coplanar waveguide resonator with a resonant frequency $\omega_{0}$.}
\end{figure}

Our model system comprises an electromagnetic resonator capacitively coupled to a single level QD as shown in Fig. \ref{fig:taba301}. The QD is coupled to a superconducting and a normal metal electrode. The total Hamiltonian of this system is given by 
\begin{equation}
\mathcal{H}(t)=\mathcal{H}_{D}+\mathcal{H}_{S}+\mathcal{H}_{N}+\mathcal{H}_{T}+\mathcal{H}_{C}(t)+\mathcal{H}_{I}(t).\label{eq:1}
\end{equation}

The first term in Eq.(\ref{eq:1}) describes the QD, which is assumed to have a single level with energy $\varepsilon_{d}$, and its Hamiltonian is given by 
\begin{align}
\mathcal{H}_{D}= & \sum_{\sigma}\varepsilon_{d}n_{\sigma}+Un_{\uparrow}n_{\downarrow},\label{eq:2}
\end{align}
where $n_{\sigma}=d_{\sigma}^{\dagger}d_{\sigma}$ is the QD's electron number operator with spin $\sigma=\uparrow,\downarrow$ and $U$ is the onsite Coulomb interaction energy in the QD. We emphasize that the single-level assumption for the QD is relevant in the situations where the broadening of the energy levels in the QD due to coupling to the electrodes and the thermal energy in the system are much smaller than the energy difference between the discrete levels in the QD \cite{PhysRevLett.65.771,PhysRevLett.66.3048}.

The second term in Eq.(\ref{eq:1}), $\mathcal{H}_{S}$, is the Hamiltonian of the superconducting electrode, which is given by 
\begin{equation}
\mathcal{H}_{S}=\underset{k,\sigma}{\sum}\varepsilon_{S,k}c_{S,k\sigma}^{\dagger}c_{S,k\sigma}+\underset{k}{\sum}\Delta(c_{S,k\uparrow}^{\dagger}c_{S,k\downarrow}^{\dagger}+h.c),\label{eq:3}
\end{equation}
where $\Delta$ is the superconducting order parameter. We assume the chemical potential of the superconducting electrode to be zero energy. In addition, the Hamiltonian of the normal metal electrode in Eq.(\ref{eq:1}) is:
\begin{equation}
\mathcal{H}_{N}=\underset{k,\sigma}{\sum}(\varepsilon_{N,k}+V_{B})c_{N,k\sigma}^{\dagger}c_{N,k\sigma},
\end{equation}
where $V_{B}$ is the bias voltage on the normal metal. Furthermore, the tunneling between electrodes and the QD is described in the total Hamiltonian by 
\begin{equation}
\mathcal{H}_{T}=\sum_{\alpha,k,\sigma}t_{\alpha}(c_{\alpha,k\sigma}^{\dagger}d_{\sigma}+h.c.),\label{eq:4}
\end{equation}
where $\alpha=N,S$ denotes the normal metal and superconducting electrodes, respectively, and $t_{\sigma}$ is the QD hybridization energy with the corresponding electrode. Also, we consider that the QD is capacitively coupled to a single-mode resonator with frequency $\omega_{0}$, which is described by the Hamiltonian 
\begin{equation}
\mathcal{H}_{C}(t)=\hbar\omega_{0}(n_p+\frac{1}{2})-i\sqrt{\gamma}\hbar E_{0}\cos(\omega_{0}t)(a-a^{\dagger}),\label{eq:5}
\end{equation}
where $n_p=a^\dagger a$ is the number operator for the photons in the resonator, $\gamma$ is the photon decay rate, and $E_{0}$ is the amplitude of an external drive on the resonator. Finally, the QD-resonator coupling is described by 
\begin{equation}
\mathcal{H}_{I}=-i\hbar\lambda(a-a^{\dagger})\left(n_{\uparrow}+n_{\downarrow}\right),\label{eq:6}
\end{equation}
where $\lambda$ is the strength of the capacitive coupling between the QD and the resonator mode. We employ mean-field approximation to decouple the operators in Eq.(\ref{eq:6}). For the resonator's operators, this approximation is analogous to the semi-classical approximation, where the single mode within the resonator is considered to exhibit classical-field like behavior. Consequently, the photon annihilation operator $a$ can be replaced with its corresponding expectation value, denoted as $\langle a\rangle=A(t)e^{-i\phi(t)}e^{-i\omega_{0}t}$, where $A(t)$ and $\phi(t)$ are the amplitude and phase of the photon mode in the resonator, respectively. Using this approximation the Hamiltonian $\mathcal{H}_{I}$ in Eq.(\ref{eq:6}) can be rewritten as 
\begin{align}
\mathcal{H}_{I}(t)= & ~2\hbar\lambda{\rm Im}[A(t)e^{-i\phi(t)}e^{-i\omega_{0}t}]\left(n_{\uparrow}+n_{\downarrow}\right)\nonumber \\
 & -i\hbar\lambda(a-a^{\dagger})\left(\langle n_{\uparrow}\rangle+\langle n_{\downarrow}\rangle\right).\label{eq:MF}
\end{align}

\subsection{\label{sec:level3}Equation of motion for the resonator mode}

In the Heisenberg representation, the equation of motion for $\left\langle a(t)\right\rangle $ reads
\begin{align}
i\frac{d}{dt}\left\langle a(t)\right\rangle = & ~\omega_{0}\left\langle a(t)\right\rangle +i\lambda\left\langle n_{\uparrow}(t)+n_{\downarrow}(t)\right\rangle \nonumber \\
 & +i\gamma\left\langle a(t)\right\rangle +i\sqrt{\gamma}E_{0}\cos(\omega_{0}t).\label{eq:11}
\end{align}
By using $\langle a\rangle=A(t)e^{-i\phi(t)}e^{-i\omega_{0}t}$, and assuming $dA/dt=d\phi/dt=0$ \cite{book:183575}, the steady state solution of the above equation determines the average photon number ($\langle n_{ph}\rangle\approx A^{2}$), and the phase ($\phi$) of the photon mode in the resonator. In order to find the steady state solution of $A$ and $\phi$, we separate the real and imaginary parts of Eq.(\ref{eq:11}) and take their time-average. We then obtain 
\begin{gather}
{\rm Re}\left[\lambda e^{i\phi}\left\langle e^{i\omega_{0}t}(n_{\uparrow}(t)+n_{\downarrow}(t))\right\rangle _{t}\right]=\gamma A-\sqrt{\gamma/2}E_{0}\cos(\phi),\label{eq:15}
\end{gather}
and 
\begin{align}
{\rm Im}\left[\lambda e^{i\phi}\left\langle e^{i\omega_{0}t}(n_{\uparrow}(t)+n_{\downarrow}(t))\right\rangle _{t}\right]=\sqrt{\gamma/2}E_{0}\sin(\phi),\label{eq:16}
\end{align}
where $\left\langle \ldots\right\rangle_{t}$ denotes time-averaging over a time period of oscillations $T=2\pi/\omega_{0}$, and is defined by
\begin{equation}
\left\langle O(t)\right\rangle _{t}=\frac{1}{T}\int_{-T/2}^{T/2}dt\langle O(t)\rangle.\label{eq:timeav}
\end{equation}

The system of equations in Eqs.(\ref{eq:15}-\ref{eq:16}) has three unknown variables $A$, $\phi$ and $\left\langle e^{i\omega_{0}t}n_\sigma(t)\right\rangle_{t}$. So, in order to find a nontrivial solution for these variable, we still need another complementary equation. At this point, we may consider the linear response of the QD in the presence of coupling to the resonator. However, as we have discussed in Appendix \ref{sec:AppLinear-response-regime}, the linear response regime is only able to demonstrate the single-photon absorption and emission processes, and it is not reliable in strong coupling regimes where nonlinear effects dominate. So, it is desirable to move beyond linear response and calculate the expectation value $\left\langle e^{i\omega_{0}t}n_\sigma(t)\right\rangle_{t}$ in the nonlinear regime. 

In the following, we will employ the Floquet-NEGF method to calculate the expectation value $\left\langle e^{i\omega_{0}t}n_\sigma(t)\right\rangle_{t}$. This way, all higher-order photon-assisted tunneling processes in the QD are non-perturbatively included in our calculations.

\subsection{\label{sec:NEGF}Floquet-NEGF formalism for QD subsystem}

As we saw in Eq.(\ref{eq:MF}), the coupling to the resonator renormalizes the QD's energy level to $\varepsilon_{d}\rightarrow\varepsilon_{d}+2\hbar\lambda A\sin(\omega_{0}t+\phi)$, resulting in a harmonic modulation of the QD's energy level. The dynamics of a quantum dot with a harmonic time dependence in its energy level, coupled to a normal metal and a superconducting electrode, can be conveniently described by the Floquet-NEGF method\cite{sun1999photon}. The advantage of this method lies in its exactness, as both the QD-leads and QD-resonator couplings are incorporated into the Green's functions of the QD to all orders of the interaction Hamiltonians $\mathcal{H}_{T}$ and $\mathcal{H}_{I}$. 

The key point in the Floquet-NEGF formalism is to understand that in the presence of a harmonic modulation with a constant frequency $\omega_{0}$, all two-time correlation functions like $G(t,t^{\prime})$ will depend on the mean time $(t+t^{\prime})/2$ only through different harmonics of the fundamental frequency $\omega_{0}$\cite{Cuevas1996}. The periodicity in the mean time allows us to represent the Fourier transform of $G(t,t^{\prime})$ in the Floquet representation as\cite{Dolcini2008,Tsuji2008}
\begin{equation}
G\left(t,t^{\prime}\right)=\sum_{m,n}\int_{-\frac{\omega_{0}}{2}}^{\frac{\omega_{0}}{2}}\frac{d\omega}{2\pi}e^{-i(\omega+m\omega_{0})t}e^{i(\omega+n\omega_{0})t^{\prime}}G_{m,n}(\omega),\label{eq:18}
\end{equation}
where $(m,n)$ correspond to various Floquet components of the transformed function.
The central quantities we need in the forthcoming calculations are the retarded and lesser Green's functions of the QD, which are defined in the Nambu basis $\Psi^{\dagger}=(d_{\uparrow}^{\dagger},d_{\downarrow})$ by $G^{R}(t,t^{\prime})=-i\theta(t-t^{\prime})\langle{\Psi(t),\Psi^{\dagger}(t^{\prime})}\rangle$ and $G^{<}(t,t^{\prime})=i\langle\Psi^{\dagger}(t^{\prime})\Psi(t)\rangle$, respectively. The other two Green's functions (advanced and greater) can be obtained in the frequency domain from the relations $G^{A}=[G^{R}]^{\dagger}$ and $G^{>}=G^{R}-G^{A}+G^{<}$, respectively.

In the Floquet representation, the Floquet components of the retarded Green's function can be obtained from the corresponding element of a large matrix containing all Floquet components of the retarded Green's function. This matrix can be calculated using Dyson's equation in the Floquet-Nambu basis, which is given by~\cite{PhysRevB.107.035410}
\begin{gather}
G^{R}\left(\omega\right)=\left\{ g^{0R}(\omega)^{-1}-\Sigma^{R}(\omega)\right\} ^{-1},\label{eq:23}
\end{gather}
where $g^{0R}$ and $\Sigma^{R}$ are the bare retarded Green's function and the  retarded self-energy of the QD, respectively. Their explicit expressions are given in Appendix \ref{app:NEGF}. In Eq.(\ref{eq:23}), all quantities are square matrices in the Floquet-Nambu basis with order $2(2N_{F}+1)$, where $N_{F}$ is the cutoff dimension for the Floquet space \cite{PhysRevB.56.1213,Zazunov2006}.

Using the retarded and advanced Green's functions obtained above, the Floquet components of the lesser Green's function can be calculated using the Keldysh equation in the Floquet-Nambu space by~\cite{PhysRevB.107.035410}
\begin{equation}
G^{<}(\omega)=G^{R}(\omega)\Sigma^{<}(\omega)G^{A}(\omega),\label{eq:25}
\end{equation}
where $\Sigma^{<}$ is the lesser self-energy of QD which its expression is given in Appendix \ref{app:NEGF}. With the Green's functions of the QD, the time-averaged occupation of the QD can be calculated as follows (see Appendix \ref{app:curr-heat} for the derivation)
\begin{gather}
\left\langle n_{\sigma}\right\rangle _{t}=\delta_{\sigma,\downarrow}-i\underset{m}{\sum}\int_{-\frac{\omega_{0}}{2}}^{\frac{\omega_{0}}{2}}\frac{d\omega}{2\pi}\left[\tau_{z}G_{m,m}^{<}\left(\omega\right)\right]_{\sigma\sigma},\label{eq:281}
\end{gather}
and 
\begin{gather}
\left\langle e^{i\omega_{0}t}n_{\sigma}(t)\right\rangle _{t}=-i\underset{m}{\sum}\int_{-\frac{\omega_{0}}{2}}^{\frac{\omega_{0}}{2}}\frac{d\omega}{2\pi}\left[\tau_z G_{m+1,m}^{<}\left(\omega\right)\right]_{\sigma\sigma}.\label{eq:282}
\end{gather}
Moreover, the time-averaged charge current through QD into electrode $\alpha=N,S$ can be calculated by~\cite{PhysRevB.74.155307,PhysRevB.107.035410}
\begin{align}
\langle I_{\alpha}(t)\rangle_{t}= & \frac{e}{2\hbar}\underset{m}{\sum}\int_{-\frac{\omega_{0}}{2}}^{\frac{\omega_{0}}{2}}\frac{d\omega}{2\pi}{\rm Tr}[\tau_{z}(G_{mm}^{R}(\omega)\Sigma_{\alpha,mm}^{<}(\omega)\nonumber \\
 & +G_{mm}^{<}(\omega)\Sigma_{\alpha,mm}^{A}(\omega)-\Sigma_{\alpha,mm}^{<}(\omega)G_{mm}^{A}(\omega)\nonumber \\
 & -\Sigma_{\alpha,mm}^{R}(\omega)G_{mm}^{<}(\omega))].\label{eq:29}
\end{align}
Similarly, the heat flux from electrode $\alpha$ into QD is given by~\cite{arrachea2013nonequilibrium,Wang2014} 
\begin{align}
\langle & J_{\alpha}(t)\rangle_{t}=  \frac{1}{2\hbar}\underset{m}{\sum}\int_{-\frac{\omega_{0}}{2}}^{\frac{\omega_{0}}{2}}\frac{d\omega}{2\pi}(\omega+m\omega_0)\nonumber \\
 & \times {\rm Tr}[\tau_{z}(G_{mm}^{R}(\omega)\Sigma_{\alpha,mm}^{<}(\omega) + G_{mm}^{<}(\omega)\Sigma_{\alpha,mm}^{A}(\omega)\nonumber \\
 & -\Sigma_{\alpha,mm}^{<}(\omega)G_{mm}^{A}(\omega)-\Sigma_{\alpha,mm}^{R}(\omega)G_{mm}^{<}(\omega))].\label{eq:31}
\end{align}

\section{Numerical results}

\subsection{\label{sec:level4}Photon absorption and emission in the resonator}

In order to assess the resonator refrigeration, we investigate the steady state of the resonator in the presence of its coupling to the QD. We assume the resonator is initially populated by an external pump, resulting in an average photon number in the resonator equal to $\langle n_p\rangle_{th}$, in the absence of coupling to the QD. Accordingly, resonator cooling can be identified when the average photon number in the resonator becomes smaller than $\langle n_p\rangle_{th}$. On the other hand, photon emission in the resonator can be deduced whenever the average photon number in the resonator becomes larger than $\langle n_p\rangle_{th}$.

Our numerical calculations are performed using a self-consistent procedure (See Appendix \ref{sec:numerical}). We have used $N_{F}=10$ for the Floquet space cut-off dimension. 
Moreover, the parameters which we have used in the numerical calculations are chosen to be near those achievable in the experimental setups \cite{lee2014spin,Bruhat-2016}. Specifically, we set $\Delta/\hbar\omega_0 = 5$, $\Gamma_S/\hbar\omega_0 = 0.01 \sim 0.1$, $\Gamma_N/\hbar\omega_0 = 0.01 \sim 0.1$, and $U = 3\Delta$ to characterize the QD system. For the resonator system, we use $\gamma/\hbar\omega_0 = 10^{-4}$ for its photon damping rate and consider $\langle n_p\rangle_{th}=20$ average photons present in the resonator when the coupling to the QD is turned off. Also, for the QD-resonator coupling strength, we take $\lambda/\hbar\omega_{0}=0.01\sim0.05$.

\begin{figure}
\includegraphics[width=1\columnwidth]{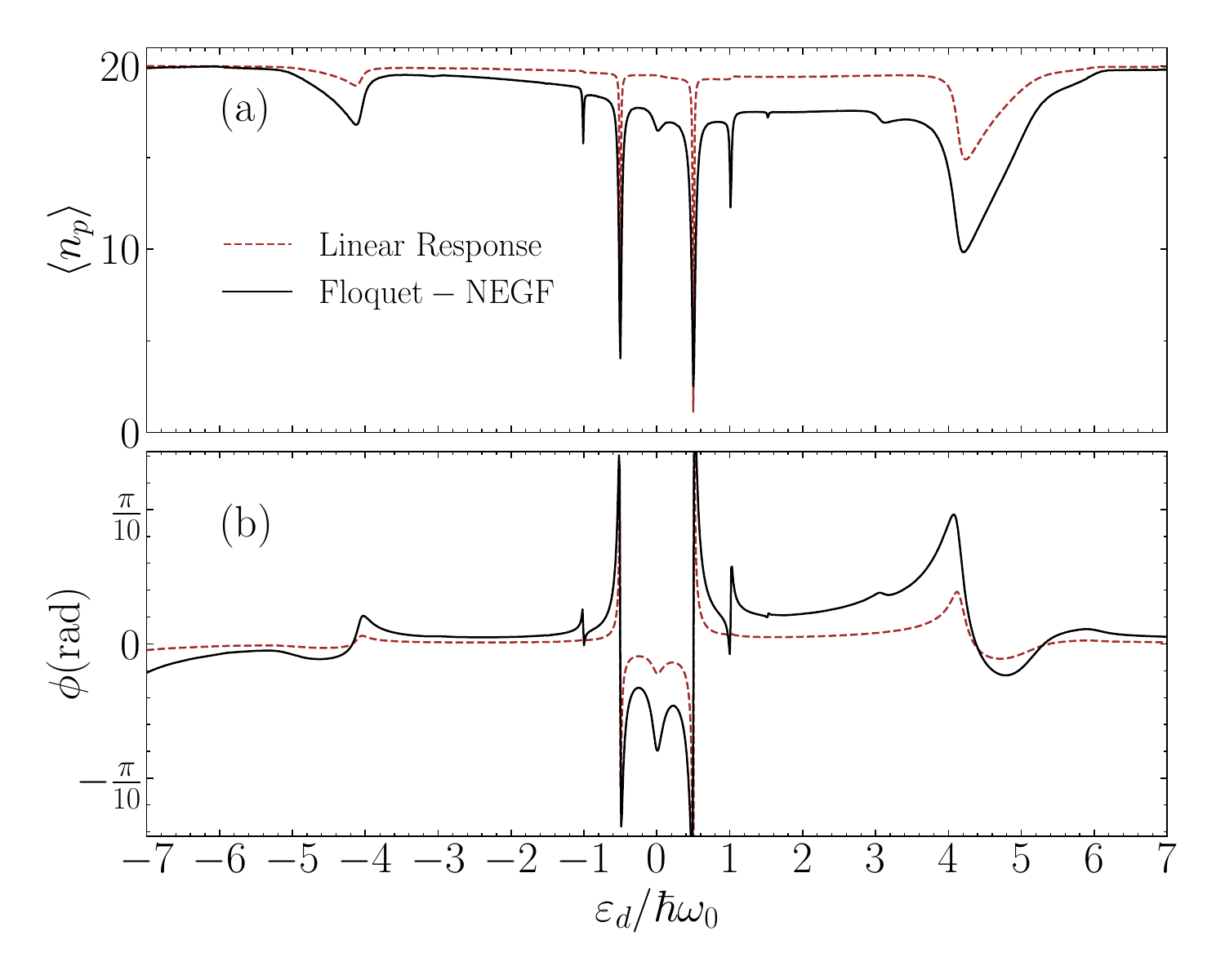} \caption{\label{fig:linear} (a) Average photon number and (b) the phase of photons in the resonator as a function of $\varepsilon_{d}$ calculated using linear response method (dashed lines) and Floquet-NEGF (solid lines) for $T_{N}=1.5\hbar\omega_{0}/k_{B}$, and $T_{S}=0$. Other parameters are $V_{B}=0$, $\Delta=5\hbar\omega_{0}$, $U=3\Delta$, $\Gamma_{S}=0.1\hbar\omega_{0}$, $\Gamma_{N}=0.01\hbar\omega_{0}$, $\gamma=10^{-4}\omega_{0}$, $\lambda=0.05\omega_{0}$, $\langle n_p\rangle_{th}=20$ and $N_F=10$. The linear response results are obtained using Eqs.(\ref{Alinearresponse}-\ref{philinearresponse}).}
\end{figure}

Numerical results for the average number of photons and their phase shift in the presence of a temperature gradient $T_{S}=0$ and $k_{B}T_{N}=1.5\hbar\omega_{0}$ are shown in Fig.\ref{fig:linear}. These quantities are easily measurable in a typical circuit-QED experiment. The solid lines in Fig.\ref{fig:linear} show the average photon number and its phase shift calculated using the Floquet-NEGF method. It is seen that the average photon number shown in Fig.\ref{fig:linear}(a), exhibits cooling peaks at different gate voltages. As we have discussed in Appendix \ref{sec:AppLinear-response-regime}, a theoretical calculation in the linear response regime allows us to find out that the sharp cooling peaks at the gate voltages $\varepsilon_{d}\approx\pm\frac{1}{2}\sqrt{(n\hbar\omega_{0})^{2}-\Gamma_{S}^{2}}$ for $n=1,2,3$ are due to photon absorptions mediated by the Andreev reflections in the QD. In Fig.\ref{fig:linear}(a), there are also a number of broad cooling peaks at gate voltages around $|\varepsilon_{d}|\approx\Delta-n\hbar\omega_{0}$, which are indeed mediated by quasiparticles that absorb some photons and tunnel through the QD into to the superconductor or vice versa. This mechanism for photon-assisted tunneling is experimentally observed in Ref.\cite{Bruhat-2016}. The linear response analysis in Appendix \ref{sec:AppLinear-response-regime} also helps us to identify the nature of the small cooling peak at $\varepsilon_{d}\approx0$. This cooling peak is mainly due to the photon absorptions caused by quasiparticles at the Fermi energy tunneling from the normal metal into the QD's level. These quasiparticles absorb a number of photons and then tunnel back to the normal metal~\cite{Bruhat-2016}. Importantly, as we show later, this type of resonator cooling does not contribute to a finite charge current through the QD. These charge fluctuations result in an effective damping of the resonator photons, which scales with $\log[\Gamma_{N}/\hbar\omega_{0}]\Gamma_{N}$ for $\Gamma_{N}\ll\hbar\omega_{0}$.

It is important to note that the weights of cooling peaks in Fig.\ref{fig:linear} are asymmetric about $\varepsilon_{d}=0$. This is due to the dependence of QD's charge susceptibility on the QD's average occupation resulting from Coulomb interaction. Particularly, at low temperatures, for $\varepsilon_{d}>0$, the QD's level is almost empty ($\langle n\rangle\approx0$), resulting in a large charge susceptibility for the QD, while for $-U<\varepsilon_{d}<0$, the QD is in the Coulomb blockade regime, where it is mostly half-filled ($\langle n\rangle\approx0.5$), giving rise to smaller values for $\chi^{R}(\omega_{0})$ (by a factor $\sim1/8$). The behavior of the phase shift of the photon mode in the resonator is shown in Fig.\ref{fig:linear}(b). We see that photon absorptions due to Andreev reflections are accompanied by strong sign changes in the phase of the photons, although the phase shifts for photon absorptions induced by quasiparticle tunnelings are not so severe.

\begin{figure}
\includegraphics[width=1\columnwidth]{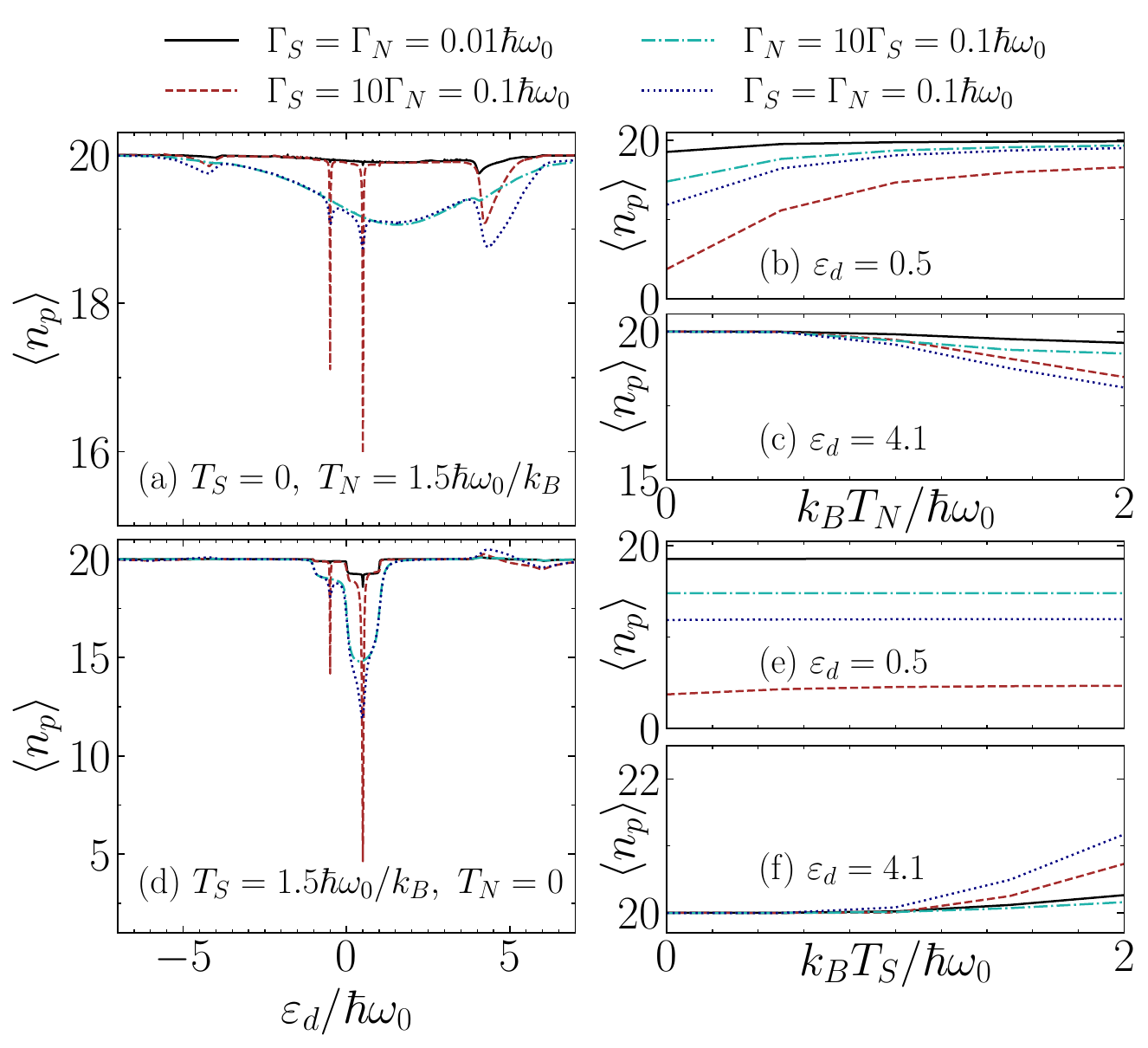} \caption{\label{fig:npnonlinear} Average photon number for $\Gamma_{S}=\Gamma_{N}=0.01\hbar\omega_{0}$ (solid), $\Gamma_{S}=10\Gamma_{N}=0.1\hbar\omega_{0}$ (dash), $\Gamma_{N}=10\Gamma_{S}=0.1\hbar\omega_{0}$ (dash-dot) and $\Gamma_{S}=\Gamma_{N}=0.1\hbar\omega_{0}$ (dot), as a function of $\varepsilon_{d}$ at $T_{N}=1.5\hbar\omega_{0}/k_{B}$, $T_{S}=0$ in (a) and at $T_{S}=1.5\hbar\omega_{0}/k_{B}$, $T_{N}=0$ in (d). Panels (b) and (c) show the photon number as a function of $T_{N}$ at $\varepsilon_{d}=0.5$ and $4.1$, respectively. Panels (e) and (f) show the photon number as a function of $T_{S}$ at $\varepsilon_{d}=0.5$ and $4.1$, respectively. Other parameters are $\Delta=5\hbar\omega_{0}$, $U=3\Delta$, $V_{B}=0$, $\gamma=10^{-4}\omega_{0}$, and $\lambda=0.01\omega_{0}$.} 
\end{figure}

Now, we investigate how the strength of coupling between the QD and electrodes, as well as the different temperature gradients over the QD, can influence the average photon number in the resonator. Fig.\ref{fig:npnonlinear}(a) shows the average photon number for different values of $\Gamma_{N}$ and $\Gamma_{S}$, when the normal metal has a higher temperature than the superconductor, i.e., $k_{B}T_{N}=1.5\hbar\omega_{0}$ and $T_{S}=0$. These results show that the Andreev cooling peaks at the central values for $\varepsilon_{d}$ are strongly suppressed for small values of $\Gamma_{S}$ (see solid and dash-dot lines in Fig.\ref{fig:npnonlinear}(a)). On the other hand, large values for $\Gamma_{N}$ result in broadening the cooling peaks, which are evident in Fig.\ref{fig:npnonlinear}(a) for $\Gamma_{N}=0.1\hbar\omega_{0}$ (see dash-dot and dotted lines).

It is also interesting to determine how the cooling peaks in Fig.\ref{fig:npnonlinear}(a) evolve as a function of normal metal temperature. Figs.\ref{fig:npnonlinear}(b)-(c) show the average photon number as a function of $T_{N}$ for two gate voltages, $\varepsilon_{d}=0.5\hbar\omega_{0}$ and $\varepsilon_{d}=4.1\hbar\omega_{0}$, respectively. As seen in Fig.\ref{fig:npnonlinear}(b), the Andreev cooling peaks have maximum weight at $T_{N}=T_{S}=0$, and become ineffective as the temperature of the normal metal is increased. Conversely, the quasiparticle cooling peaks are absent at zero temperature, and become visible only after the thermal energy of thermally excited quasiparticles becomes comparable to the continuum of the superconductor (see Fig.\ref{fig:npnonlinear}(c)).

In Fig.\ref{fig:npnonlinear}(d), we assume that the superconductor has a higher temperature than the normal metal with $k_{B}T_{S}=1.5\hbar\omega_{0}$ and $T_{N}=0$, and study the resonator cooling as a function of $\varepsilon_{d}$. Here, we see that for small values of $\Gamma_{N}=\Gamma_{S}=0.01\omega_{0}$, only a small Andreev cooling peak survives at $\varepsilon_{d}=0.5\hbar\omega_{0}$, while for a larger $\Gamma_{N}$ value, this peak becomes broader (See solid and dash-dot lines in Fig.\ref{fig:npnonlinear}(d)). For larger values of $\Gamma_{S}$, the Andreev cooling peaks become very strong, but the quasiparticle peak at $\varepsilon_{d}=4.1$ shows emission instead of absorption. This emission peak can be easily understood by noting that for $T_{S}>T_{N}$, there are some thermally excited quasiparticles above the gap edge, which can emit a photon into the resonator and tunnel into the QD's level when its energy is about $\hbar\omega_{0}$ below the gap edge (This corresponds to $\varepsilon_{d}=4.1\hbar\omega_{0}$ in our case). By the same reasoning, we can understand the cooling peak at $\varepsilon_{d}=6\hbar\omega_{0}$, which is approximately $\hbar\omega_{0}$ above the gap edge and favors photon absorption-assisted quasiparticle tunneling from the superconductor into the QD's level.

The temperature dependence of the Andreev cooling peak (at $\varepsilon_{d}=0.5\hbar\omega_{0}$) and the quasiparticle emission (at $\varepsilon_{d}=4.1\hbar\omega_{0}$) are shown in Figs.\ref{fig:npnonlinear}(e)-(f). As expected, the Andreev cooling peak is not largely affected by the temperature of the superconductor because the Cooper pairs in the subgap region are not influenced by the superconductor temperature. On the other hand, the quasiparticle emission peaks are absent at zero temperature and start to show up only at sufficiently large values of $T_{S}$.

\begin{figure}
\includegraphics[width=1\columnwidth]{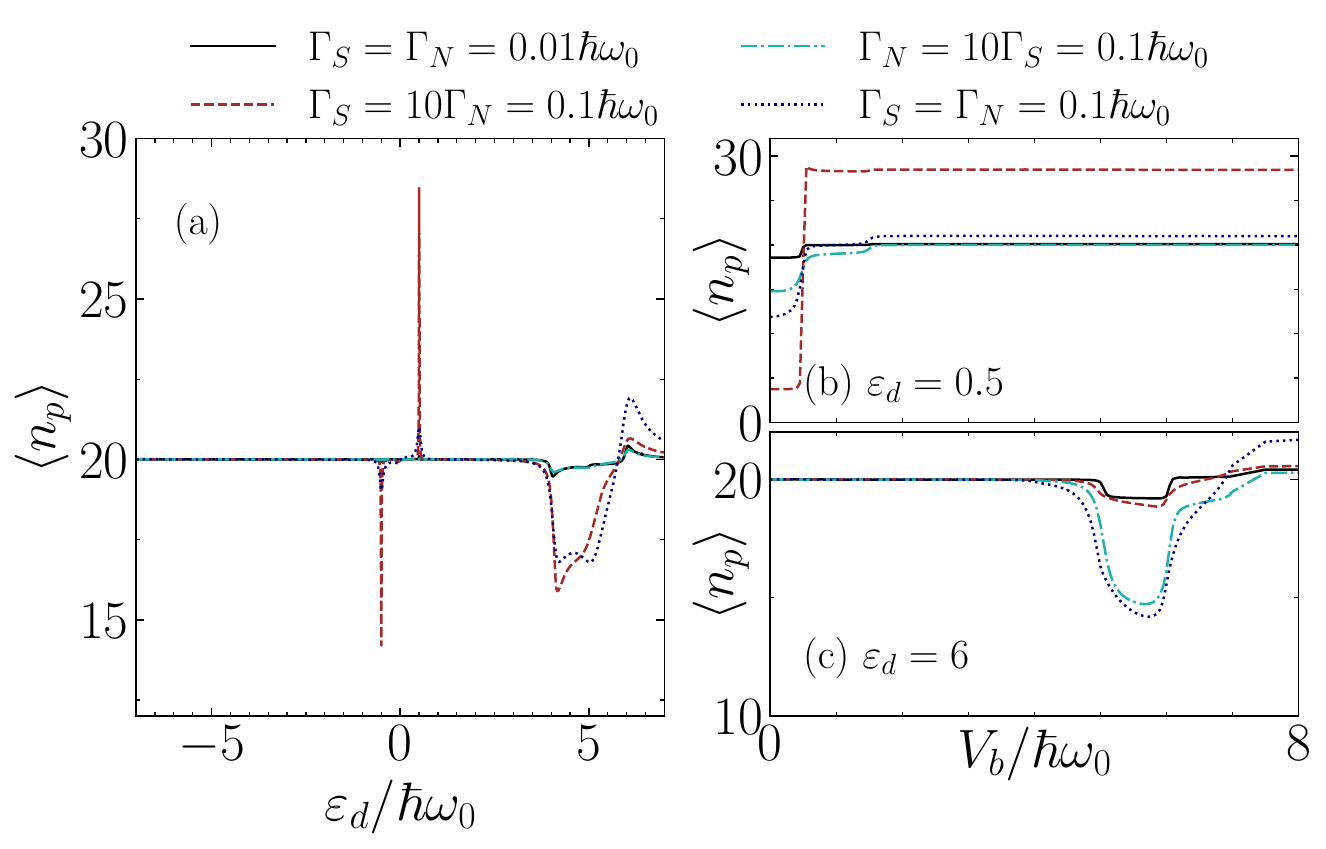}\caption{\label{fig:npnonlinear12}(a) Average photon number for $\Gamma_{S}=\Gamma_{N}=0.01\hbar\omega_{0}$ (solid), $\Gamma_{S}=10\Gamma_{N}=0.1\hbar\omega_{0}$ (dash), $\Gamma_{N}=10\Gamma_{S}=0.1\hbar\omega_{0}$ (dash-dot) and $\Gamma_{S}=\Gamma_{N}=0.1\hbar\omega_{0}$ (dot), as a function of $\varepsilon_{d}$ at $V_{B}=10\hbar\omega_{0}$. Panels (b) and (c) show the photon number as a function of $V_{B}$ at $\varepsilon_{d}=0.5$ and $6$, respectively. Other parameters are $\Delta=5\hbar\omega_{0}$, $U=3\Delta$, $T_{S}=T_{N}=0$, $\gamma=10^{-4}\omega_{0}$, and $\lambda=0.01\omega_{0}$ .}
\end{figure}

Having clarified the behavior of the resonator-QD coupled system in the presence of a finite temperature gradient over the QD, we now investigate our model system when a finite voltage bias is applied to the normal metal. Fig. \ref{fig:npnonlinear12}(a) shows the Floquet-NEGF results when a large voltage bias $V_{B}=10\hbar\omega_{0}$ is applied to the normal metal. Here, we assume that both electrodes have the same temperatures ($T_{N}=T_{S}=0$) to ensure that thermoelectric effects do not affect the results. In general, photon-assisted Andreev reflection in an QD in the presence of a finite bias may lead to either photon absorption or emission, depending on the value of $\varepsilon_{d}$ \cite{sun1999photon}. This behavior is evident in Fig. \ref{fig:npnonlinear12}(a), where we observe resonator cooling at $\varepsilon_{d}=-0.5\hbar\omega_{0}$, while strong photon emission is present at $\varepsilon_{d}=0.5\hbar\omega_{0}$. On the other hand, around the gap edge where photon-assisted quasiparticle tunnelings are more probable, we observe a combination of photon absorption at $\Delta-\hbar\omega_{0}<\varepsilon<\Delta$ and photon emission for $\varepsilon>\Delta$.

The dependence of these photon absorptions and emissions on different values of $\Gamma_{N}$ and $\Gamma_{S}$ has also been studied in Fig. \ref{fig:npnonlinear12}(a). It is observed that for small values of $\Gamma_{S}$ ($\Gamma_{S}=0.01$), the Andreev peaks are strongly suppressed, while the quasiparticle absorptions and emissions are still present around the gap edge. By increasing the value of $\Gamma_{S}$, the Andreev and quasiparticle peaks become more pronounced. Especially, the Andreev peaks are stronger for $\Gamma_{S}>\Gamma_{N}$, while the quasiparticle emission peaks become stronger with increasing $\Gamma_{N}$. We emphasize that the model parameters for the experiment in Ref. \cite{Bruhat-2016} are very similar to the case of $\Gamma_{S}\ll\Gamma_{N}$ in Fig. \ref{fig:npnonlinear12}(a), where both quasiparticle photon absorptions and emissions are present around the gap edge, while the Andreev contributions are absent.

The voltage dependence of the photon-assisted Andreev and quasiparticle peaks at $\varepsilon_{d}=0.5\hbar\omega_{0}$ and $\varepsilon_{d}=6\hbar\omega_{0}$ is shown in Figs. \ref{fig:npnonlinear12}(b)-(c), respectively. As we saw previously, for $\Gamma_{S}=0.1\hbar\omega_{0}$, the Andreev absorption peaks are present even at $V_{B}=T_{N}=T_{S}=0$. The results in Fig. \ref{fig:npnonlinear12}(b) show that this behavior persists for small bias voltages unless the voltage bias reaches a threshold value ($V_{B}\approx0.5\hbar\omega_{0}$), above which the absorption peak transforms into an emission peak. The situation for the quasiparticle peak at $\varepsilon_{d}=6\hbar\omega_{0}$, as shown in Fig. \ref{fig:npnonlinear12}(c), is somewhat different. At small bias voltages, there are no photon absorptions. When the bias voltage reaches $V_{B}\approx\varepsilon_{d}-\hbar\omega_{0}$, quasiparticle tunneling in the QD favors photon absorption. For larger bias voltages ($V_{B}>\varepsilon_{d}$), photon emissions in the resonator become dominant.

\subsection{\label{sec:level5}Charge Current and Heat Fluxes}

In the previous section, we investigated cooling and photon emission in the resonator due to the capacitive coupling to the QD. From a thermodynamic point of view, the heat flux taken from the resonator should be injected into the QD system. We note that while charge conservation establishes $I\equiv I_{N}=-I_{S}$, the heat fluxes must obey the energy conservation rule as 
\begin{gather}
IV_{b}=-\underset{\alpha\in N,S}{\sum}J_{\alpha}-\dot{Q}_{ph},\label{eq:conserve}
\end{gather}
where $\dot{Q}_{ph}$ is the heat flux from the resonator. Therefore, in our model system when we assumed zero bias voltage $V_{b}=0$, the cooling power of the resonator is given by 
\begin{gather}
P_{cool}\equiv\dot{Q}_{ph}=-\underset{\alpha\in N,S}{\sum}J_{\alpha}.\label{eq:conserve1}
\end{gather}

\begin{figure}
\includegraphics[width=1\columnwidth]{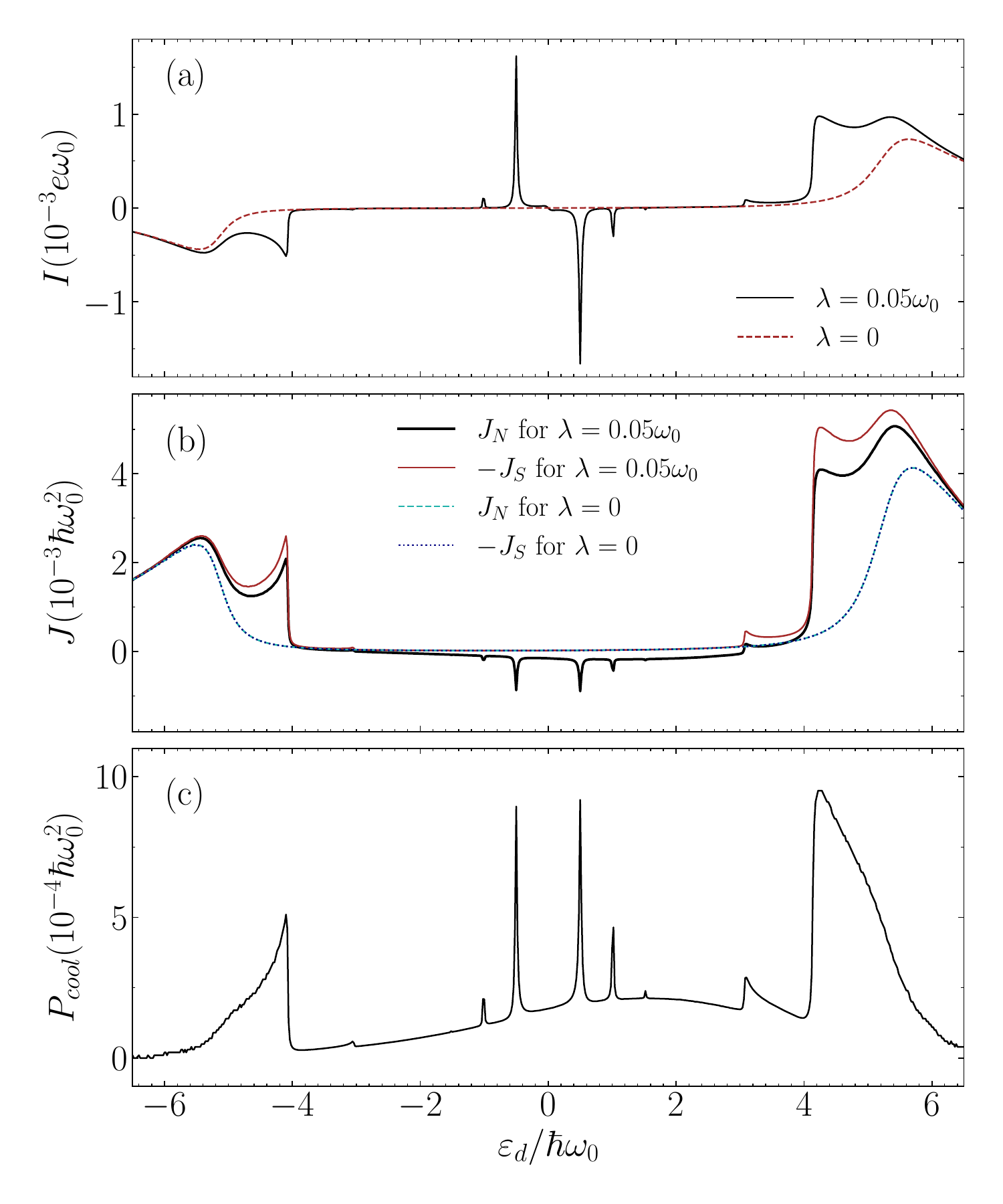} \caption{\label{fig3}(a) Charge current ($I$), (b) heat currents in both terminals ($J_{N}$ and $J_{S}$), and (c) cooling power of resonator ($P_{cool}$) as functions of $\varepsilon_{d}$ for $\Gamma_{S}=10\Gamma_{N}=0.1\hbar\omega_{0}$, $T_{S}=0$ and $T_{N}=1.5\hbar\omega_{0}/k_{B}$ at $\lambda=0$ and $\lambda=0.05\omega_{0}$. Other parameters are as in Fig.\ref{fig:linear}}
\end{figure}

In Fig. \ref{fig3}, we consider $V_{B}=0$ and show the charge current, heat flux, and the cooling power of the resonator as functions of $\varepsilon_{d}$ for the cases with $\lambda=0$ and $\lambda=0.05\omega_{0}$, with $k_{B}T_{N}=1.5\hbar\omega_{0}$ and $T_{S}=0$, respectively. When the QD is decoupled from the resonator ($\lambda=0$), the only contribution to the thermoelectric current in Fig. \ref{fig3}(a) is due to quasiparticle tunneling, where thermally excited quasiparticles can tunnel from the normal metal into the continuum of the superconductor through the QD's level. These tunneling events are maximal around $\varepsilon_{d}\approx\pm\Delta$. As expected, the charge current changes sign depending on the value of $\varepsilon_{d}$ \cite{Verma_2022}.

By turning on the coupling between the QD and the resonator, new transport channels in the QD become available, originating from photon-assisted electron tunneling through the QD. This is shown in Fig.\ref{fig3}(a) for $\lambda=0.05\omega_{0}$. Note that the behavior of the average photon number in the resonator for this parameter configuration was shown in Fig.\ref{fig:npnonlinear12}(c). 
Importantly, the sign of the current enhancements is determined by both the photon absorption mechanism that enhances the current and the value of $\varepsilon_{d}$. Focusing on the gate voltages around $\varepsilon_{d}=0$, where photon absorptions give rise to Andreev reflections in the QD, we observe in Fig.\ref{fig3}(a) that for $\varepsilon_{d}/\hbar\omega_{0}=0.5,~1$ and $1.5$, current enhancements have a negative sign, indicating that charge current flows from the superconductor into the normal metal, while for $\varepsilon_{d}/\hbar\omega_{0}=-0.5$ and $-1$, current enhancements are in the opposite direction. This behavior of photon absorption-induced charge current in the Andreev regime can be interpreted by noting that, for example, when $\varepsilon_{d}=\hbar\omega_{0}/2$, a Cooper pair from the superconductor is broken into two electrons. These electrons can perform an Andreev reflection by absorbing one photon from the resonator and injecting two electrons into the normal metal, resulting in a negative charge current in Fig.\ref{fig3}(a).

A different mechanism is responsible for the charge current enhancements around the superconducting gap edges at $|\varepsilon_{d}|>\Delta-n\omega_{0}$ in Fig.\ref{fig3}(a). Here, quasiparticles in the QD's level can absorb photons from the resonator to reach energies around the BCS peak and tunnel into the superconductor, resulting in a charge current with a positive sign. Therefore, the maximum enhancement of the quasiparticle current occurs at $|\varepsilon_{d}|\approx\Delta-n\omega_{0}$.

It is worth mentioning that the resonator cooling regions around $\varepsilon_{d}=0$ in Fig. \ref{fig:npnonlinear12} do not accompany any current enhancements in Fig. \ref{fig3}. This is because resonator coolings at these regions are due to quasiparticles that tunnel from the normal metal into the QD's level, absorb a number of photons from the resonator, and then tunnel back into the normal metal without giving rise to a finite charge current between the electrodes.

The heat fluxes in both terminals are depicted in Fig. \ref{fig3}(b). For the case of $\lambda=0$, a finite positive heat current is present at the gate voltages around the superconducting gap edges. Here, thermally excited quasiparticles tunnel between the normal metal and the superconductor, carrying heat from the normal metal with higher temperature toward the superconductor with lower temperature. Thus, for $\lambda=0$, as implied by the energy conservation rule for $T_{N}>T_{S}$, which we considered here, the thermoelectric current through the QD always cools down the normal metal ($J_{N}>0$), while the superconductor is heated up by the same heat flux ($J_{S}=-J_{N}<0$) \cite{Verma_2022}.

For $\lambda\neq0$, photon absorption-assisted tunneling can enhance the heat currents. As shown in Fig. \ref{fig3}(b), normal metal heat currents mediated by photon-assisted quasiparticle tunneling have a positive sign irrespective of the gate voltage values. On the other hand, the superconductor is heated up by a larger heat current, given by $-J_{S}=J_{N}+P_{cool}$, where $P_{cool}$ is the cooling power of the resonator, provided in Eq. (\ref{eq:conserve1}) and shown in Fig. \ref{fig3}(c). Interestingly, for gate voltages in the Andreev regime, the superconducting heat current is zero ($J_{S}=0$) because no heat currents can be carried by the Cooper pairs. Thus, in the Andreev regime, all heat fluxes coming from the resonator are consumed by the normal metal, resulting in a negative sign for the normal metal heat flux ($J_{N}<0$). This implies that in the Andreev regime, the resonator can function as a heat pump, injecting heat into the normal metal irrespective of the temperature difference between the two electrodes.

\section{\label{sec:level6}CONCLUSIONS}


We have studied the photon absorption and emission in an electromagnetic resonator which is coupled to a hybrid quantum dot system. We have shown that in the weak coupling limit, our results are consistent with the linear response regime where the photon assisted tunneling processes in QD are included up to the lowest order of interaction between the resonator and the QD. In the nonlinear regime, multiphoton processes are included in the calculations by employing the Floquet-NEGF method. Using this formalism, we have studied the coupled QD-resonator system in the presence of either a voltage bias or a temperature gradient over the QD.

We have found that two main photon-assisted charge tunneling mechanisms in the QD can be responsible for photon absorption or emission in the resonator. First, the Andreev mechanism manifests when the QD's level energy is deep in the subgap region, positioned around a point where a quasiparticle can absorb or emit a number of photons to complete the Andreev reflection process. We have shown that this mechanism is dominant when $\Gamma_{S}$ is large enough and the level broadening due to coupling to the normal metal is small. Additionally, it is found that this mechanism can lead to large resonator coolings even under equilibrium conditions with zero voltage and thermal biases. Furthermore, we have observed that Andreev reflections can result in photon emissions in the resonator when the QD is voltage-biased.

The second photon-assisted tunneling mechanism is attributed to quasiparticle tunnelings, which are predominantly observed when the QD's energy level is near the gap edge. We have demonstrated that in the presence of a finite voltage bias, this mechanism can result in both photon absorption and emission, contingent upon the energy of the QD's level. Furthermore, under a thermal bias, it is illustrated that the QD can either absorb or emit photons in the resonator, depending on the value of $\varepsilon_{d}$ and the direction of the thermal bias. Additionally, we have investigated the dependence of these photon absorption and emission processes on the voltage and thermal bias.

We have also investigated the charge current in the QD and the heat fluxes between the resonator and the electrodes. It is demonstrated that photon absorption and emissions in the resonator can enhance both the charge current and the heat fluxes. This suggests that the heat flux of either part of the system can be controlled by tuning the QD's energy level, as well as the voltage and thermal bias over the QD.

Our calculations rely on a combination of the Floquet-NEGF method and semiclassical laser equations. The semiclassical laser equations govern the dynamics of the photon mode in the resonator, while the Floquet-NEGF method enables us to calculate the response of the QD when it is coupled to the resonator. One advantage of using the Floquet-NEGF method is its ability to properly include all higher-order photon-assisted electron tunnelings between the QD's level and the electrodes in the calculations. Particularly, this approach allows for the nonperturbative treatment of the finite energy gap of the superconducting electrode. 

It is worth to mention that an alternative method for calculating the multiphoton response of QD is the Floquet master equation formalism, as described in Ref.\cite{kohler2018dispersive}. However, in this approach, the coupling between the QD and electrodes is treated perturbatively, and thus the finite energy gap of the superconductor cannot be exactly accounted for, as we achieve using Dyson's equation in the NEGF formalism.


\begin{acknowledgments}
We thank Alireza Bahrampour for useful comments. N.J. acknowledges the support from Research Center for Quantum Engineering and Photonics Technology, Sharif University of Technology.
\end{acknowledgments}

\appendix

\section{Expressions for $G^{R}(\omega)$ and $\Sigma^{R,<}(\omega)$}

\label{app:NEGF}

The retarded Green's function of QD in Eq.(\ref{eq:23}) is given by~\cite{PhysRevB.107.035410}
\begin{equation}
G^{R}\left(\omega\right)=\left\{ g^{0R}(\omega)^{-1}-\Sigma^{R}(\omega)\right\} ^{-1}.\label{eq:app1}
\end{equation}
Here, $g^{0R}(\omega)$ is the matrix representing the bare retarded Green's function of the isolated QD in the infinite-$U$ limit~\cite{haug2008quantum}. Its matrix elements, given by $2\times2$ block matrices in the Nambu space, are as follows: 
\begin{equation}
g_{mn}^{0R}(\omega)=\delta_{mn}\frac{1-\langle n_{\sigma}\rangle}{\omega_{m}+i\eta-\varepsilon_{d}\tau_{z}},\label{eq:g0rappA}
\end{equation}
where, $\omega_{m}=\omega+m\omega_{0}$ for $m\in\mathbb{Z}$, $\delta_{mn}$ is the Kronecker delta and $\tau_{z}$ is the third Pauli matrix in the Nambu space. Moreover, in Eq.(\ref{eq:app1}), $\Sigma^{R}(\omega)$ represents the retarded self-energy, given by
\begin{equation}
\Sigma_{mn}^{R}(\omega)=\Pi_{mn}^{R}+\delta_{mn}(\Sigma_{S}^{R}(\omega_{m})+\Sigma_{N}^{R}(\omega_{m})).\label{eq:app2}
\end{equation}
Here, $\Pi^{R}$ is the selfenergy that accounts for the harmonic modulation of the QD's energy level, and its matrix elements are given by 
\begin{equation}
\Pi_{mn}^{R}=-i\lambda A\left(e^{-i\phi}\delta_{m,n+1}-e^{i\phi}\delta_{m,n-1}\right)\tau_{z}.\label{eq:24}
\end{equation}
Moreover, in Eq.(\ref{eq:app2}), the terms $\Sigma_{S}^{R}(\omega)$ and $\Sigma_{N}^{R}(\omega)$ are the self-energies that account for the coupling between the QD and the electrodes, and their expressions are given respectively by
\cite{trocha2014spin}
\begin{equation}
\Sigma_{S}^{R}(\omega)=-i\Gamma_{S}\beta(\omega)(I+\frac{\Delta}{\hbar\omega}\tau_{x}),\label{eq:21}
\end{equation}
and 
\begin{equation}
\Sigma_{N}^{R}(\omega)=-i\Gamma_{N}I,\label{eq:22}
\end{equation}
where $I$ is the $2\times2$ unit matrix, and the parameter $\beta(\omega)$, which is related to the normalized BCS density of states, is given by $\beta(\omega)=\frac{\left|\omega\right|}{\sqrt{\omega^{2}-\Delta^{2}}}\theta\left(\left|\hbar\omega\right|-\Delta\right)-i\frac{\omega}{\sqrt{\Delta^{2}-\omega^{2}}}\theta\left(\Delta-\left|\hbar\omega\right|\right)$. We use the wide-band approximation, where the hybridization of the QD's energy level with the electrodes takes the simple form $\Gamma_{N,S}\equiv\pi|t_{N,S}|^{2}\rho_{0}^{N,S}$, and $\rho_{0}^{N}$ and $\rho_{0}^{S}$ are the frequency-independent densities of states of the normal lead and the normal state of the superconducting electrode, respectively.

We emphasize that the method used here to consider the effects of both the Coulomb interaction and the coupling to the electrodes in the Green's function of the QD is known as the Hubbard-I approximation. Within this approximation, the Coulomb interaction in the QD is taken into account exactly, while its coupling to the electrodes is considered through a decoupling scheme in the equation of motion for the Green's functions \cite{Lim2020,Verma2024}. 

Now, the lesser Green's function of QD can be calculated from the relation~\cite{PhysRevB.107.035410}
\begin{equation}
G^{<}(\omega)=G^{R}(\omega)\Sigma^{<}(\omega)G^{A}(\omega),\label{eq:gless}
\end{equation}
where, $\Sigma^{<}(\omega)$ is the lesser self-energy which is obtained by using the Ng ansatz \cite{Ng1996} as
\begin{equation}
\Sigma^{<}(\omega)=\sum_{\alpha=S,N}(\Sigma_{\alpha}^{A}(\omega)-\Sigma_{\alpha}^{R}(\omega))f_{\alpha}(\omega-\tau_{z}\mu_{\alpha}).
\end{equation}
Here, $f_{\alpha}(\omega)$ is the Fermi-Dirac distribution function for the superconducting and normal metal electrodes, which is given by $f_{S(N)}\left(\omega\right)=(1+\exp[\hbar\omega/k_{B}T_{S(N)}])^{-1}$, and $T_{\alpha\in{N,S}}$ denotes the temperature in electrode $\alpha$.

\section{\label{sec:AppLinear-response-regime}Linear response regime}

We start with the calculation of $\left\langle e^{i\omega_{0}t}n(t)\right\rangle _{t}$ within the linear response regime. In order to calculate the linear response of the QD's average occupation, one can expand $G^{<}$ to the linear order in the QD-resonator coupling term in Eq.(\ref{eq:MF}), which gives 
\begin{equation}
G^{<}(t,t)=g^{<}(t,t)-i\lambda(Ae^{-i\phi}e^{-i\omega_{0}t}-h.c.)\chi^{R}(\omega_{0}),\label{eq:glesLR}
\end{equation}
where $g^{<}(t,t^\prime)$ is the lesser Green's function of QD calculated from Eq.(\ref{eq:gless}) for $\lambda=0$, and $\chi^{R}(\omega)$ is the Fourier transform of the retarded charge susceptibility of the QD, defined as $\chi^{R}(t-t^{\prime})=-i\theta(t-t^{\prime})\langle[n(t),n(t^{\prime})]\rangle_{\lambda=0}$. Here, $\langle...\rangle_{\lambda=0}$ denotes averaging with respect to the state of the QD when it is decoupled from the resonator~\cite{PhysRevB.93.075425}. Using analytical continuation rules for contour-ordered correlation functions \cite{haug2008quantum}, the expression for $\chi^{R}(\omega)$ is obtained as 
\begin{align}
\chi^{R}(\omega) & =\int d\omega_{1}{\rm Tr}[g^{<}(\omega_{1})\nonumber \\
 & \times\tau_{z}\left(g^{R}(\omega_{1}-\omega)+g^{A}(\omega_{1}+\omega)\right)\tau_{z}].\label{chiRexp}
\end{align}

Now, multiplying Eq.(\ref{eq:glesLR}) by $e^{i\omega_{0}t}$, and taking its time-average, we obtain 
\begin{equation}
\left\langle e^{i\omega_{0}t}n(t)\right\rangle _{t}=\lambda Ae^{-i\phi}\chi^{R}(\omega_{0}),\label{n_av_lin}
\end{equation}
Then, by substituting Eq.(\ref{n_av_lin}) into Eqs.(\ref{eq:15}-\ref{eq:16}), we find the steady-state solutions of the amplitude and phase of the photon mode in the resonator as
\begin{gather}
A=\frac{\sqrt{\gamma}E_{0}}{2\left|\lambda^{2}\chi^{R}(\omega_{0})+\gamma\right|},\label{Alinearresponse}\\
\phi=\arccos\left[\frac{\lambda^{2}{\rm Re}[\chi^{R}(\omega_{0})]+\gamma}{\left|\lambda^{2}\chi^{R}(\omega_{0})+\gamma\right|}\right].\label{philinearresponse}
\end{gather}

The integration required for calculating $\chi^{R}(\omega_{0})$ is complicated in the general case. However, it can be analytically solved in some limiting cases. For example, by assuming $\Gamma_{S}\approx0$ and $U\rightarrow\infty$, we find that the expression for $\chi^{R}(\omega_{0})$ at zero temperature and bias voltage is given by (see Appendix \ref{sec:AppD})
\begin{equation}
\chi^{R}(\omega_{0})=\frac{(1-\langle n\rangle)^{3}\Gamma_{N}}{\pi\omega_{0}(2\Gamma_{N}+i\omega_{0})}\log\left[\frac{\varepsilon_{d}^{2}+\Gamma_{N}^{2}}{\varepsilon_{d}^{2}+(\Gamma_{N}+i\omega_{0})^{2}}\right].\label{chiRUinfty}
\end{equation}
This shows that even when the QD is in equilibrium and is effectively coupled only to a normal metal electrode, the quasi-particle fluctuations between the normal metal and the QD's level around the Fermi energy can give rise to a finite damping of the photon mode in the resonator \cite{Bruhat-2016}, which scales with $\log[\Gamma_{N}/\hbar\omega_{0}]\Gamma_{N}$ for $\Gamma_{N}\ll\hbar\omega_{0}$. In addition, Eq.(\ref{chiRUinfty}) reveals that the charge susceptibility of the QD depends on the average occupation of the QD through the term $(1-\langle n\rangle)^{3}$, which is due to the Coulomb interaction in the QD.

Another limiting configuration for calculating $\chi^{R}(\omega_{0})$ is when $\Gamma_{S}$ is not negligible, but the superconducting energy gap is large ($\Delta\rightarrow\infty$). In this case, the integration in $\chi^{R}(\omega_{0})$ can be solved analytically for $T=V_{b}=0$ and $U\rightarrow\infty$, resulting in a lengthy expression which we do not present here. However, it is worth mentioning that from this expression, we find that the real part of $\chi^{R}(\omega_{0})$ shows large enhancements at two subgap gate voltages $\varepsilon_{d}=\pm\frac{1}{2}\sqrt{(\hbar\omega_{0})^{2}-\Gamma_{S}^{2}}$, indicating the possibility for resonator cooling at these gate voltages. In fact, cooling at these gate voltages is mediated by electron tunnelings in the QD that need to absorb one photon to complete the process of Andreev reflection in the QD\cite{sun1999photon}.

The numerical results obtained from Eqs.(\ref{Alinearresponse}-\ref{philinearresponse}) are shown by dashed lines in Fig.\ref{fig:linear} in the main text. It is seen that only the single photon cooling peaks due to either the quasiparticle or the Andreev reflection-assisted photon absorption are shown in the linear regime, while the Floquet-NEGF method captures all higher-order cooling peaks because of its nonperturbative nature.

\section{\label{sec:AppD}Calculation of $\chi^{R}(\omega)$ in the linear response regime}

When the QD is decoupled from the resonator and is only coupled to a normal electrode and in the limit of large Coulomb interaction ($U\rightarrow\infty$), the retarded, advanced and lesser Green's functions of the QD are given by (assuming $\langle n\rangle=\langle n_{\uparrow}\rangle=\langle n_{\downarrow}\rangle$)
\begin{align}
g^{R}(\omega) & =[g^{A}(\omega)]^\dagger=\left([g^{0R}(\omega)]^{-1}+i\Gamma_{N}\right)^{-1}\nonumber \\
 & =\frac{1-\langle n\rangle}{\omega-\varepsilon_{d}+i(1-\langle n\rangle)\Gamma_{N}},\label{eq:gRappC}\\
g^{<}(\omega) & =\frac{2i(1-\langle n\rangle)^{2}\Gamma_{N}}{(\omega-\varepsilon_{d})^{2}+(1-\langle n\rangle)^{2}\Gamma_{N}^{2}}f(\omega).
\end{align}
Note that the large Coulomb interaction limit is considered in Eq.(\ref{eq:gRappC}), by substituting $g^{0R}(\omega)$ with the expression given in Eq.(\ref{eq:g0rappA}). 

Now, the integration in Eq.(\ref{chiRexp}) can be written as
\begin{align}
\chi^{R}(\omega)=\int_{-\infty}^{0} & d\omega_{1}\frac{2i(1-\langle n\rangle)^{2}\Gamma_{N}}{(\omega_{1}-\varepsilon_{d})^{2}+(1-\langle n\rangle)^{2}\Gamma_{N}^{2}}\nonumber \\
 & \times\left(\frac{(1-\langle n\rangle)}{\omega_{1}-\omega-\varepsilon_{d}+i(1-\langle n\rangle)\Gamma_{N}}\right.\nonumber \\
 & \left.\,\,+\frac{(1-\langle n\rangle)}{\omega_{1}+\omega-\varepsilon_{d}-i(1-\langle n\rangle)\Gamma_{N}}\right).
\end{align}
This integration can be easily solved, giving the final result in Eq.(\ref{chiRUinfty}).

\section{Calculation of time-averaged quantities using Floquet-NEGF formalism}

\label{app:curr-heat}

We first derive Eq.(\ref{eq:281}) of the main text. The average electron occupation with spin $\sigma$ in the QD is given by $\langle n_{\sigma}(t)\rangle=\delta_{\sigma,\downarrow}-i[\tau_{z}G^{<}(t,t)]_{\sigma\sigma}$, where $G^{<}$ is the lesser Green's function of the QD in the Nambu basis. Then, the time-averaged occupation of the QD is calculated by
\begin{align}
\left\langle n_{\sigma}\right\rangle _{t} & =\frac{1}{T}\int_{-T/2}^{T/2}dt\left\langle n_{\sigma}(t)\right\rangle \nonumber \\
 & =\delta_{\sigma,\downarrow}-i\frac{1}{T}\int_{-T/2}^{T/2}dt[\tau_{z}G^{<}(t,t)]_{\sigma\sigma}\nonumber \\
 & =\delta_{\sigma,\downarrow}-i\underset{m}{\sum}\int_{-\frac{\omega_{0}}{2}}^{\frac{\omega_{0}}{2}}\frac{d\omega}{2\pi}\left[\tau_{z}G_{m,m}^{<}\left(\omega\right)\right]_{\sigma\sigma},\label{eq:281-1}
\end{align}
where the last line is obtained by using Eq.(\ref{eq:18}) and employing the relation
\begin{equation}
\frac{1}{T}\int_{-T/2}^{T/2}dte^{i(m-n)\omega_{0}t}=\delta_{mn}.
\end{equation}
In the same way, by noting that $\frac{1}{T}\int_{-T/2}^{T/2}dte^{i\omega_{0}t}\delta_{\sigma,\downarrow}=0$ and $\frac{1}{T}\int_{-T/2}^{T/2}dte^{-i(m-n-1)\omega_{0}t}=\delta_{m,n+1}$, we can obtain the expression given in Eq.(\ref{eq:282}) for $\left\langle e^{i\omega_{0}t}n_{\sigma}(t)\right\rangle _{t}$. 

Also, the time-averaged charge current from QD into electrode $\alpha=N,S$ can be written in the Nambu basis as~\cite{haug2008quantum}
\begin{align}
\left\langle I_{\alpha}\right\rangle _{t}= & \frac{e}{2\hbar}\frac{1}{T}\int_{-T/2}^{T/2}dt\int_{-\infty}^{\infty}dt^{\prime}{\rm Tr}[\tau_{z}\nonumber \\
 & (G^{R}(t,t^{\prime})\Sigma_{\alpha}^{<}(t^{\prime},t)+G^{<}(t,t^{\prime})\Sigma_{\alpha}^{A}(t^{\prime},t)\nonumber \\
 & -\Sigma_{\alpha}^{<}(t,t^{\prime})G^{A}(t^{\prime},t)-\Sigma_{\alpha}^{R}(t,t^{\prime})G^{<}(t^{\prime},t))].\label{eq:currapp}
\end{align}
By transforming Eq.(\ref{eq:currapp}) to the Floquet representation, we reach to Eq.(\ref{eq:29}) in the main text for the charge current.
\section{\label{sec:numerical}Numerical implementation}

We saw that Eqs.(\ref{eq:15}, \ref{eq:16}, \ref{eq:281} and \ref{eq:282}) form a closed system of equations that must be solved self-consistently. Our implementation to solve these equations is as follows: 

We start with initial guesses $A=\sqrt{\langle n_p\rangle_{th}},\,\phi=0,$ and $\left\langle n_{\sigma}\right\rangle =0.5$, where $\langle n_p\rangle_{th}=E_0^2/2\gamma$ is the average photon number in the resonator in the absence of resonator-QD coupling. Then, we calculate the lesser Green's function $G^{<}$ in Eq.(\ref{eq:25}). From this, we can calculate $\langle e^{i\omega_{0}t}n_{\sigma}(t)\rangle$ using Eq.(\ref{eq:282}). Then, the new values for $A$ and $\phi$ can be calculated by employing Eqs.(\ref{eq:15}-\ref{eq:16}). Subsequently, we calculate the new value for $\langle n_{\sigma}\rangle$ using Eq.(\ref{eq:281}). We then compare the new values of these quantities with their previous ones and repeat this procedure until convergence is reached. In our calculations, the difference criteria needed for stopping the iterations was set to $10^{-5}$.

\bibliography{Myreference}

\end{document}